\documentclass[12pt,preprint]{aastex}

\shortauthors{Sil'chenko}
\shorttitle{Lenticular Galaxies}

\received{2005 May 16}
\begin{document}

\title{Stellar populations in nearby lenticular galaxies}
\slugcomment{Based on the observations collected with the 6m
telescope (BTA) at the Special Astrophysical Observatory (SAO) of the
Russian Academy of Sciences (RAS).}

\author{O. K. Sil'chenko}
\affil{Sternberg Astronomical Institute, Moscow, 119992 Russia\\
       and Isaac Newton Institute of Chile, Moscow Branch\\
     Electronic mail: olga@sai.msu.su}

\begin{abstract}
We have obtained 2D spectral data for a sample of 58 nearby S0 galaxies
with the Multi-Pupil Spectrograph of the 6m telescope of the Special
Astrophysical Observatory of the Russian Academy of Sciences. The Lick 
indices H$\beta$, Mgb, and $\langle \mbox{Fe} \rangle$ are calculated separately 
for the nuclei and for the bulges taken as the rings between $R=4\arcsec$ and 
$7\arcsec$; and the luminosity-weighted ages, metallicities, and Mg/Fe ratios 
of the stellar populations are estimated by confronting the data to SSP models.
Four types of galaxy environments are considered: clusters, centers
of groups, other places in groups, and field. The nuclei are found to be 
on average slightly younger than the bulges in any types of environments, 
and the bulges of S0s in sparse environments are younger than those in dense 
environments. The effect can be partly attributed to the well-known age
correlation with the stellar velocity dispersion in early-type galaxies
(in our sample the galaxies in sparse environements are in average less 
massive than those in dense environments), but for the most massive S0s,
with $\sigma _*=170-220$ km/s, the age dependence on the environment 
is still significant at the confidence level of 1.5$\sigma$.
\end{abstract}


\section{Introduction}

In classical morphological sequence by \citet{hubble} lenticular
galaxies occupy intermediate position between ellipticals and spirals:
they have a smooth and red appearance as the ellipticals, but also have
stellar disks, almost as large as those of the spirals. The most popular
hypothesis of S0 origin is that of their transformation from the spirals
by stopping global star formation and removing or consuming remaining
gas \citep{ltcs0}. In distant, $z\sim 0.5$, clusters this 
transformation is now observed directly: the number of lenticulars in the clusters
diminishes strongly with the redshift \citep{fasano}, instead one can see
`passive spirals' -- red spiral galaxies lacking star formation --
at the periphery (`infalling regions') of the intermediate-redshift clusters
\citep{goto, yamauchi}. Many theoretical works have been done to explain in detail 
what physical mechanisms may be involved into the process of spiral 
transformation into the lenticulars: tidally induced collisions of disk gas clouds
\citep{byrdval90}, harassment \citep{moore96}, ram pressure by intercluster 
medium \citep{quilis}, etc. For the S0s in the field, the scheme of their 
transformation from the spirals is not so clear, but common view is that some 
external action like minor merger may produce the necessary effect.

By reviewing the various mechanisms of secular evolution which may transform 
a spiral galaxy into a lenticular one we have noticed that most of them result
in gas concentration in the very center of the galaxy, so that a nuclear star
formation burst seems inavoidable circumstance of the S0 galaxy birth.
If to refer to S0 statistics in the clusters located between $z=0$
and $z\approx 1$, the main epoch of S0 formation is $z\approx 0.4-0.5$,
so the nuclear star formation bursts in the nearby S0s must not be older
than 5 Gyr. Indeed, in my spectral study of the central parts of nearby
galaxies in different types of environments \citep{me93} I have found
that $\sim 50\%$ of nearby lenticulars have strong absorption lines
H$\gamma$ and H$\delta$ in their nuclear spectra so they are of `E+A' type,
as it is presently called and are dominated by intermediate-age stellar
population. In this respect the S0s have resembled rather early-type spirals
than ellipticals. Here I aim to continue this study, with a larger sample
and with panoramic spectral data in order to separate the nuclei and 
their outskirts (bulges) which is a substantial advantage with respect to 
aperture spectroscopy.

Another crucial point of the present study, and also of a global paradigm
of galaxy formation, is environmental influence. The current hierarchical
assembly paradigm predicts a younger age of galaxies in lower density
environments -- for the most recent simulations see e.g. \citet{lanzoni05}
or \citet{delucia05}. Observational evidences concerning early-type galaxies are 
controversial: some authors find differences of stellar 
population ages between the clusters and the field 
\citep{terlfor,kunt,thogal},
some authors do not find any dependence of the stellar population age 
on environment density \citep{gallens}. In order to check whether the mean
ages of the stellar populations depend on environment density monotonously,
as the hierarchical paradigm predicts, in this work I consider four types of 
environments separately: the cluster galaxies, the brightest (central)
galaxies of groups, the second-ranked group members, and the field galaxies.

\section{Sample}

The sample of lenticular galaxies considered in this work consists of 
58 objects, mostly nearby and bright. It does not pretend to be complete but 
rather representative. In the LEDA we have found 122 galaxies in total with
the following parameters: $-3 \le T \le 0$, $v_r<3000$ km/s, 
$B_T^0 <13.0$, $\delta _{2000.0} >0$, without bright AGN or intense
present star formation in a nucleus; among them 40 Virgo members.
For our sample, from this list we have selected 8 Virgo members and 
42 other galaxies --  half of the rest. A few galaxies are added to broaden
the luminosity range: NGC 5574, NGC 3065, and NGC 7280 are fainter 
than $B_T^0=13.0$, NGC 80 and NGC 2911 are very luminous but farther from us 
than 40 Mpc. 

\clearpage

\begin{deluxetable}{llcccccl}
\tabletypesize{\scriptsize}
\rotate
\tablenum{1}
\tablewidth{24cm}
\tablecaption{Our sample of S0 galaxies\label{sample}}
\tablehead{
\colhead{Galaxy\tablenotemark{a}}  &  
\colhead{Environment\tablenotemark{b}} &
\colhead{Type\tablenotemark{c}}  &
\colhead{$\sigma _0$, km/s\tablenotemark{d}}  &
\colhead{$v_r$, km/s\tablenotemark{d}}  &
\colhead{Dates (green)} &
\colhead{Dates (red)} &
\colhead{Detailed description, ref.}
}
\startdata
 N0080 & Group center & SA0- &  260 & 5698 & Aug96, Oct03 & -- & 
 \citet{n80} \nl
 N0474 & Pair & (R$^{\prime}$)SA(s)$0^0$ &
 164 & 2372 & Oct03 & -- & -- \nl
 N0524 & Group center & SA(rs)0+ &  253 & 2379 & Oct97 &
  Oct96 & \citet{faceon} \nl
 N0676 & Pair & S0/a & $140^e$ & 1506 & 
 Oct03 & -- & -- \nl
 N0936 & Group center & SB(rs)0+ & 190 & 1430 & Oct02, Oct03 &
  Oct02 & -- \nl
 N1023 & Group center & SB(rs)0- & 204 & 637 & Oct96 & -- & 
 \citet{lense1} \nl
 N1161 & Pair & S0 & $185^e$ & 1954 & Oct03 & -- & -- \nl
 N2300 & Group member & SA0+ & 261 & 1938 & Sep01 & -- & -- \nl
 N2549 & Group center & SA(r)0+ &  143 & 1039 & Oct04 & Oct02 & -- \nl
 N2655 & Group center & SAB(s)0/a & 163 & 1404 & Oct99, Oct00 & Oct00 &
 \citet{ourortho} \nl
 N2681 & Group center & (R')SAB(rs)0/a & 108 & 692 & Sep01 & Mar02 & -- \nl
 N2685 & Field & (R)SB0+pec &  94 & 883 & Oct94 & -- & \citet{polar} \nl
 N2732 & Pair & S0 & 154 & 1960 & Oct00 & Sep01 &
 \citet{ourortho} \nl
 N2768 & Group center & S0$_{1/2}$ &  182 & 1373 & Jan01 & Oct00 &
 \citet{ourortho} \nl
 N2787 & Field & SB(r)0+ &  194 & 696 & Oct00 & Oct00 &
  \citet{ourortho} \nl
 N2880 & Field & SB0- &  136 & 1608 & Sep01 & -- & -- \nl
 N2911 & Group center & SA(s)0: & 234 & 3183 & Dec99 & Jan98 &
 \citet{ourortho} \nl
 N2950 & Field & (R)SB(r)$0^0$ & 182 & 1337 & Oct03 & Oct05 & -- \nl
 N3065 & Group center & SA(r)0+ &  160 & 2000 & Sep01 & Oct05 & -- \nl
 N3098 & Field & S0 & 105 & 1311 & Jan01 & -- & -- \nl
 N3166 & Group member & SAB(rs)0/a & 112 & 1345 & Mar03 & Jan98 & -- \nl
 N3245 & Group member & SA(r)$0^0$ & 210 & 1358 & Mar03 & -- & -- \nl
 N3384 & Group member & SB(s)0- & 148 & 704 & Dec99 & -- & 
  \citet{leo1} \nl
 N3412 & Group member & SB(s)$0^0$ & 101 & 841 & Mar04 & -- & -- \nl
 N3414 & Group center & S0pec &  237 & 1414 & Jan01 & Mar02 &
  \citet{ourortho} \nl
 N3607 & Group center & SA(s)0+ &  224 & 935 & Apr01 & Mar02 & -- \nl
 N3941 & Group center & SB(s)0$^0$ & 159 & 928 & Mar03 & -- & -- \nl
 N3945 & Group member & SB(rs)0+ & 174 & 1259 & Mar03 & -- & -- \nl
 N4026 & UMa cluster & S0 & 178 & 930 & Mar03 & -- & -- \nl
 N4036 & Group center & S0- & 189 & 1445 & May97, Jan98 & Jan98 &
 \citet{n4036} \nl
 N4111 & UMa cluster & SA(r)0+ & 148 & 807 & Jan01 & Mar02 &
  \citet{ourortho} \nl
 N4125 & Group center & E6 pec & 227 & 1356 & Mar03 & Jan98 & -- \nl
 N4138 & UMa cluster & SA(r)0+ & 140 & 888 & Jan98, Dec99 & Dec99 &
 \citet{counter} \nl
 N4150 & Group member & SA(r)0+ & 85 & 226 & Apr01 & Mar02 & -- \nl
 N4179 & Group center & S0 & 157 & 1256 & Mar03 & -- & --  \nl
 N4233 & Virgo cluster & S$0^0$ & 220 & 2371 & Apr02 & Apr02 &
  \citet{ourortho} \nl
 N4350 & Virgo cluster & SA0 & 181 & 1200 & Jan01 & -- & -- \nl
 N4379 & Virgo cluster & S0- & 108 & 1069 & Jun99 & -- & -- \nl
 N4429 & Virgo cluster & SA(r)0+ & 192 & 1106 & Jun99 & May97 &
  \citet{rings} \nl
 N4526 & Virgo cluster & SAB(s)0+ & 264 & 448 & Apr01 & Mar02 & -- \nl
 N4550 & Virgo cluster & SB0 & 91 & 381 & Jan98, Jun99 & Jun99 &
  \citet{counter} \nl
 N4570 & Virgo cluster & S0/E7 & 188 & 1730 & Mar04 & -- & -- \nl
 N4638 & Virgo cluster & S0- & 122 & 1164 & Mar04 & -- & -- \nl
 N4866 & Pair & SA(r)0+ &  210 & 1988 & Apr01 & -- & -- \nl
 N5308 & Group member & S0- & 211 & 2041 & Mar03 & -- & -- \nl
 N5422 & Group member & S0 & 165 & 1820 & Mar03 & -- & -- \nl
 N5574 & Group member & SB0-? & 75 & 1659 & Jun99 & -- & 
   \citet{lense2} \nl
 N5866 & Group member & S$0_3$ & 159 & 672 & Aug98 & May96 & -- \nl
 N6340 & Group center & SA(s)0/a & 144 & 1198 & Aug96, Oct97 & Aug96 &
  \citet{faceon} \nl
 N6548 & Pair & SB0 & $121^e$ & 2179 & Oct04 & -- & -- \nl
 N6654 & Pair & (R')SB(s)0/a & 149$^e$ & 1821 & Sep01 
  & -- & -- \nl
 N6703 & Field & SA0- & 180 & 2461 & Oct03 & Oct03 & -- \nl
 N7013 & Field & SA(r)0/a & 84 & 779 & Oct96, Aug98 & Aug96 &
  \citet{rings} \nl
 N7280 & Field & SAB(r)0+ &  104 & 1844 & Aug98 & Oct98 &
  \citet{we7280} \nl
 N7332 & Field & S0 pec &  124 & 1172 & Aug96, Oct97  & -- & 
  \citet{lense1} \nl
 N7457 & Field & SA(rs)0-? &  69 & 812 & Oct99, Dec99 & -- &
  \citet{lense2} \nl
 N7743 & Field & (R)SB(s)0+ & 84 & 1710 & Oct03 & -- & -- \nl
 U11920 & Field & SB0/a & $116^e$ & 1145 & Oct03 & -- & -- \nl
\enddata
\tablenotetext{a}{Galaxy ID -- N=NGC, U=UGC}
\tablenotetext{b}{From Guiricin et al. 2000}
\tablenotetext{c}{Hubble type from the NED}
\tablenotetext{d}{Mainly from the LEDA}
\tablenotetext{e}{From our observations}
\end{deluxetable}

\clearpage

Table~\ref{sample} lists all the galaxies with some of their characteristics 
such as morphological type, redshift, and central velocity dispersion. Sorting 
of the galaxies according to their environment type has been made by using 
the NOG group catalogue \citep{nog}; we have only classified three galaxies
belonging to the Ursa Major cluster following \citet{tullyuma}. Our sample 
includes 11 cluster galaxies, from Virgo and Ursa Major, 17 central 
(the brightest) galaxies of groups with 3 members and more, 18 second-ranked 
group members to which we have added paired galaxies, and 12 field lenticulars
which are defined as not mentioned in the NOG catalogue at all.

All the 
galaxies of Table~\ref{sample} have been observed with the integral-field unit
-- the Multi-Pupil Fiber/Field Spectrograph (MPFS) 
\footnote{http://www.sao.ru/hq/lsfvo/devices/mpfs/mpfs\_main.html} 
\citep{mpfsref} of the 6m telescope 
of the Special Astrophysical Observatory of the Russian Academy of Sciences
between 1994 and 2005. For these years the instrument was modified 
more than once.
We started with the field of view  of $10\arcsec \times 12\arcsec$,
with the spatial element (pupil) size of $1\farcs 3$, with the spectral
resolution of 5~\AA, and spectral range less than 600~\AA. Now we have
$16\arcsec \times 16\arcsec$, 
the spatial element (pupil) size of $1\arcsec$, the spectral
resolution of 3.5~\AA, and spectral range of 1500~\AA. Usually we observe
two spectral ranges, the green one centered onto $\lambda$5000~\AA,
and the red one centered onto the H$\alpha$ line. The optical
design had been modified too: two different schemes, a TIGER-like one --
for the description of the instrumental idea of the TIGER mode of IFU
one can see \citet{betal95} -- and
that with fibers, were used before and after 1998. 
We have described in detail 23 of 58 lenticulars in our previous papers 
(see the references in the Table~\ref{sample})
where one can find not only the characteristics
of the various versions of the MPFS, but also 2D maps of Lick indices and
kinematical parameters. Here we consider only two discrete areas of every
galaxy -- the unresolved nuclei and the wide rings, with $R_{in}=4\arcsec$
and $R_{out}=7\arcsec$, which we are treating as the `bulges'. The boundaries 
of the rings have been selected as a compromise between the seeing limitation
(the seeing FWHM are typically $2\farcs 5$ at the 6m telescope)
in order to avoid the influence of the nuclei on the bulge measurements, 
and the size of our field of view 
which causes incomplete azimuthal coverage at $R>7\arcsec$. At our limit
distance, $D=40$ Mpc, the outer radius of the `bulge' areas, $7\arcsec$, 
corresponds to the linear size of 1.35 kpc. The nuclei are presented by the
integrated fluxes over the central spatial elements within the maximum radius
of 0.1 kpc from the centers. 

\clearpage

\begin{table}
\tablenum{2}
\caption{The comparison of two independent index determinations with the MPFS
\label{compint}}
\begin{tabular}{r|cc|cc|cc|}
\tableline
 NGC &  
\multicolumn{2}{|c|}{H$\beta$, \AA} & 
 \multicolumn{2}{|c|}{Mgb, \AA} & 
\multicolumn{2}{|c|}{$\langle \mbox{Fe} \rangle$, \AA} \\
 & nucleus & bulge & nucleus & bulge & nucleus & bulge \\
\tableline
80 & 1.57 & 1.70 & 5.12 & 4.44 & 3.14 & 3.22 \\
 & 1.66 & $1.59\pm 0.20$ & 5.00 & $4.44 \pm 0.04$ & 2.94 & $2.95\pm 0.15$ \\
\tableline
936 & 1.13 & $1.41\pm 0.03$ & 4.64 & $4.51\pm 0.03$ & 2.86 & $2.50\pm 0.01$ \\
 & 1.41 & $1.07\pm 0.07$ & 4.93 & $4.53 \pm 0.10$ & 3.20 & $2.80\pm 0.07$ \\
\tableline
2655 & 1.56 & $1.55\pm 0.03$ & 3.77 & $3.60\pm 0.11$ & 2.10 & $2.07\pm 0.05$ \\
 & 1.73 & $1.35\pm 0.05$ & 3.70 & $3.69 \pm 0.02$ & 2.38 & $2.47\pm 0.02$ \\
\tableline
4036 & 0.12 & $0.92\pm 0.08$ & 5.56 & $4.09\pm 0.13$ & 2.56 & $2.64\pm 0.07$ \\
 & 0.82 & $0.80\pm 0.08$ & 5.85 & $3.61 \pm 0.25$ & 3.28 & -- \\
\tableline
4138 & 1.14 & $1.10\pm 0.03$ & 4.76 & $3.34\pm 0.15$ & 2.97 & $2.00\pm 0.14$ \\
 & 0.74 & $0.96\pm 0.06$ & 4.66 & $3.45 \pm 0.21$ & 2.65 & $2.11\pm 0.07$ \\
\tableline
4550 & 1.64 & 1.92 & 3.18 & 3.13 & -- & -- \\
 & 1.64 & $1.41\pm 0.03$ & 3.20 & $3.14 \pm 0.05$ & 2.53 & $1.95\pm 0.08$ \\
\tableline
6340 & 1.05 & 0.86 & 4.65 & 3.06 & 2.92 & 2.10 \\
 & 1.56 & 1.24 & 4.49 & 3.18 & 2.76 & 2.12 \\
\tableline
7013 & 1.63 & 2.03 & 3.84 & 3.32 & 2.99 & -- \\
 & 1.58 & $2.15\pm 0.09$ & 3.78 & $3.27 \pm 0.05$ & 3.00 & $2.35\pm 0.09$ \\
\tableline
7332 & 2.10 & $1.54\pm 0.10$ & 3.67 & $2.54\pm 0.20$ & 2.92 & $2.05\pm 0.16$ \\
 & 2.24 & $1.65\pm 0.10$ & 3.80 & $2.77 \pm 0.12$ & 2.80 & 2.23 \\
\tableline
7457 & 1.93 & 2.27 & 2.72 & 3.37 & 2.49 & 2.24 \\
 & 1.99 & $2.21\pm 0.05$ & 2.92 & $2.98 \pm 0.06$ & 2.44 & $2.26\pm 0.07$ \\
\tableline

\end{tabular}
\end{table}

\clearpage

\begin{table}
\tablenum{3}
\caption{The mean differences between our indices and the Trager's et al. data
\label{compext}}
\begin{tabular}{lccc}
\tableline
 &  H$\beta$ & Mgb & $\langle \mbox{Fe} \rangle$ \\
\tableline
$\Delta$ & $+0.07$\AA & $-0.05$\AA & $+0.12$\AA \\
 & $\pm 0.06$\AA & $\pm 0.07$\AA & $\pm 0.09$\AA \\
\tableline
\end{tabular}
\tablecomments{The second line of the table contains the formal errors
of the mean offsets of our index system with respect to the Lick one}
\end{table}

\clearpage

The Lick indices H$\beta$, Mgb, Fe5270, and Fe5335 have been measured
for the nuclei and for the bulges of all the galaxies; farther we use the composite
iron index $\langle \mbox{Fe} \rangle \equiv$(Fe5270+Fe5335)/2. During all our 
observational runs we observed standard stars from \citet{woretal} and
calibrated our index system onto the standard Lick one. The measured
indices were corrected for the stellar velocity dispersions; we calculated
the corrections by artificial broadening of the spectra of the standard
stars. We estimate the typical statistical accuracy in each of three indices (defined 
by the S/N ratio which has been kept as 70-90 (per \AA) in the nuclei and
$\sim 30$ at the edges of the frames) as 0.1~\AA. Some galaxies of the
sample have been observed twice. In the Table~\ref{compint} we show
the raw index measurements from two independent observational runs for each
of those objects; $\pm$ accompanying the bulge indices reflect partly the
index variations along the radii -- we average four measurements at four $R$'s 
from $4\arcsec$ to $7\arcsec$ for each galaxy and give here the errors of the means.
The mean absolute difference between two independent index measurements 
is 0.20~\AA\ for the nuclei and 0.18~\AA\ for the bulges over the 
Table~\ref{compint}. If we analyse three indices separately, we obtain
the mean absolute differences ( the rms of the differences) of
0.22~\AA\ (0.29~\AA) for H$\beta$,
0.15~\AA\ (0.19~\AA) for Mgb, and
0.22~\AA\ (0.28~\AA) for the composite iron index. These results mean
that the accuracy of the Mgb corresponds to our expectations from the
S/N statistics, namely, is 0.1~\AA, and the accuracy of the H$\beta$ and
$\langle \mbox{Fe} \rangle$ is somewhat worse, namely, is  0.15~\AA.
Among our 58 galaxies, 28 objects have Lick index measurements
through the central aperture $2\arcsec \times 4\arcsec$ by \citet{trager98}.
The results of the comparison of these standartized Lick indices 
with our measurements for the nuclei are presented 
in Table~\ref{compext} and in Fig.~1. The smallest scatter is found for
H$\beta$ and the largest one -- for $\langle \mbox{Fe} \rangle$, that
is consistent with the fact the among the four indices, H$\beta$, Mgb, Fe5270,
and Fe5335, the errors quoted by \citet{trager98} are the smallest for
H$\beta$ (0.24~\AA\ on average over the common list) 
and the largest -- for Fe5335 (0.34~\AA\ on average over the common list).   
In general, our index system does not 
deviate from the standard Lick one in any systematic way, so we can 
determine the stellar population properties by confronting our indices 
to evolutionary synthesis models.

\clearpage

\begin{deluxetable}{llcccccccc}
\tabletypesize{\scriptsize}
\rotate
\tablenum{4}
\tablewidth{22cm}
\tablecaption{Indices and ages for the nuclei of the S0 galaxies\label{nuclei}}
\tablehead{
\colhead{Galaxy\tablenotemark{a}}  &  
\colhead{Environment\tablenotemark{b}} &
\colhead{H$\beta$}  &
\colhead{Mgb}  &
\colhead{$\langle \mbox{Fe} \rangle$} &
\colhead{T\tablenotemark{c}, Gyr} &
\colhead{[Z/H]\tablenotemark{c}} &
\colhead{EW([OIII]5007), \AA } &
\colhead{$T^{\prime}$\tablenotemark{d}, Gyr} &
\colhead{$\mbox{[Z/H]} ^{\prime}$\tablenotemark{d}}
}
\startdata
 N0080 & Group center & 1.62 &  5.06 & 3.04 & 7 & +0.4 & 0.08 & 6 & +0.5 \nl
 N0474 & Group member (pair) & 1.70 & 4.55 & 3.14 & 4 & +0.4 & 0.94 & 2 &
  $\ge +0.7$ \nl
 N0524 & Group center & 1.33 &  4.87 & 2.68 & 14 & +0.1 & 0.46 & 10 & +0.2 \nl
 N0676 & Group member (pair) & 1.02 & 4.16 & 2.90 & $> 15$ & 0 & 2.0 & 3 &
 +0.7 \nl
 N0936 & Group center & 1.27 & 4.78 & 3.03 & 15 & +0.2 & 0.79 & 5 & +0.4 \nl
 N1023 & Group center & 1.57 & 5.03 & 2.99 & 8 & +0.4 & 0.12 & 7 & +0.4 \nl
 N1161 & Group member (pair) & 1.84 & 5.31 & 3.04 & 3 & +0.7 & 0.06 &
 4 & +0.7 \nl
 N2300 & Group member & 1.64 & 5.19 & 2.87 & 7 & +0.4 & 0 & 7 & +0.4 \nl
 N2549 & Group center & 2.51 &  4.47 & 3.32 & $< 2$ & $\ge +0.7$ &
  0.20 & $< 2$ & $\ge +0.7$ \nl
 N2655 & Group center & 1.65 &  3.74 & 2.24 & 2 & 0 & 2.51 & 2 & 0 \nl
 N2681 & Group center & 3.52 & 2.31 & 2.02 & $< 2$ & $\le 0$ & 0.56
 & $< 2$ & $\le 0$ \nl
 N2685 & Field & 1.75 & 3.59 & 2.58 & 4 & +0.1 & 0.57 & 4 & +0.1 \nl
 N2732 & Group member (pair) & 1.88 & 3.55 & 2.71 & 7 & 0 & 0.46 & 3 & 
  +0.3 \nl
 N2768 & Group center & 0.91 &  4.90 & 2.64 & 11 & +0.2 & 0.91 & 15 & 
  +0.1 \nl
 N2787 & Field & 0.61 &  5.25 & 2.12 & $> 15$ & 0 & 0.95 & $> 15$ & 0 \nl
 N2880 & Field & 1.72 &  4.15 & 2.63 & 9 & +0.1 & 0.1 & 8 & +0.2 \nl
 N2911 & Group center & --0.11 & 5.65 & 2.59 & 15 & +0.1 & 2.38 & 
  $> 15$ & 0 \nl
 N2950 & Field & 2.66 & 4.67 & 3.23 & $< 2 $ & $\ge +0.7$ & 0.28 &
  $< 2 $ & $\ge +0.7$ \nl
 N3065 & Group center & 0.42 &  4.16 & 2.42 & --$^e$ & --$^e$ & 2.36 &
  6 & +0.2 \nl
 N3098 & Field & 1.79 & 3.65 & 2.20 & 10 & --0.2 & 0.28 & 7 & --0.1 \nl
 N3166 & Group member & 2.36 & 3.68 & 2.94 & $< 2$ & +0.7 & 0.57 &
  $< 2$ & +0.7 \nl
 N3245 & Group member & 0.67 & 4.52 & 2.96 & 6 & +0.3 & 0.61 &
  $ > 15$ & +0.1 \nl
 N3384 & Group member & 2.04 & 4.64 & 3.07 & 3 & +0.7 & 0.05 & 3 & +0.7 \nl
 N3412 & Group member & 2.33 & 4.00 & 3.02 & 2 & +0.7 & 0.23 & $< 2 $
  & +0.7 \nl
 N3414 & Group center & 0.82 &  5.21 & 2.74 & 13 & +0.2 & 1.23 & 7 & 
  +0.4 \nl
 N3607 & Group center & 0.93 &  5.24 & 2.78 & 12 & +0.2 & 0.71 & 15 &
  +0.2 \nl
 N3941 & Group center & 1.69 & 4.61 & 3.26 & 4 & +0.7 & 0.83 & 2 & 
  $\ge +0.7$ \nl
 N3945 & Group member & 1.44 & 4.74 & 3.28 & 6 & +0.5 & 0.30 & 7 & +0.5 \nl
 N4026 & UMa cluster & 1.73 & 4.44 & 3.11 & 6 & +0.4 & 0 & 6 & +0.4 \nl
 N4036 & Group center & 0.47 & 5.70 & 2.92 & 11 & +0.3 & 1.42 & 10 & 
  +0.4 \nl
 N4111 & UMa cluster & 1.99 & 4.60 & 2.56 & $< 2$ & +0.7 & 0.54 & 2
 &  +0.6 \nl
 N4125 & Group center & 1.31 & 4.66 & 3.14 & 7 & +0.4 & 0.70 & 5 & +0.5 \nl
 N4138 & UMa cluster & 0.94 & 4.71 & 2.81 & 12 & +0.2 & 4.7 & $< 2$
  & $\ge +0.7$ \nl
 N4150 & Group member & 2.65 & 2.51 & 1.60 & 2 & --0.2 & 0.87 & $< 2$
  & --0.1 \nl
 N4179 & Group center & 1.90 & 4.94 & 3.31 & 4 & +0.7 & 0 & 4 & +0.7 \nl
 N4233 & Virgo cluster & 1.06 & 4.80 & 3.00 & 15 & +0.2 & 0.78 & 10 &
  +0.3  \nl
 N4350 & Virgo cluster & 1.41 & 5.26 & 2.91 & 8 & +0.4 & 0.17 & 10 & 
  +0.4  \nl
 N4379 & Virgo cluster & 1.51 & 4.36 & 2.45 & 15 & 0 & 0.14 & 13 & 0 \nl
 N4429 & Virgo cluster & 1.60 & 4.61 & 2.96 & 3 & +0.7 & 0.25 & 6 & +0.4 \nl
 N4526 & Virgo cluster & 1.62 & 4.75 & 2.78 & 3 & +0.7 & 0.24 & 6 & +0.4 \nl
 N4550 & Virgo cluster & 1.64 & 3.20 & 2.53 & 5 & 0 & 1.16 & 3 & +0.2 \nl
 N4570 & Virgo cluster & 1.72 & 5.18 & 2.86 & 5 & +0.4 & 0 & 5 & +0.4 \nl
 N4638 & Virgo cluster & 2.01 & 4.75 & 3.42 & 3 & +0.7 & 0.06 & 3 & 
  +0.7 \nl
 N4866 & Group member (pair) & 1.28 & 4.60 & 2.85 & 8 & +0.3 & 0.69 &
  8 & +0.3 \nl
 N5308 & Group member & 1.48 & 5.14 & 2.92 & 11 & +0.3 & 0 & 11 & +0.3 \nl
 N5422 & Group member & 1.41 & 4.85 & 3.28 & 12 & +0.4 & 0.52 & 5 & +0.5 \nl
 N5574 & Group member & 2.78 & 2.48 & 2.47 & 2 & 0 & 0.25 & $< 2 $ & 0 \nl
 N6340 & Group center & 1.30 & 4.57 & 2.84 & 11 & +0.2 & 0.33 & 13 & +0.2 \nl
 N6548 & Group member (pair) & 1.67 & 4.58 & 2.90 & 8 & +0.3 & 0 & 8 & 
  +0.3 \nl
 N6654 & Group member (pair) & 1.67 & 4.51 & 2.78 & 8 & +0.3 & 0 & 8 & 
  +0.3 \nl
 N6703 & Field & 1.49 & 4.34 & 3.14 & 12 & +0.2 & 0.33 & 7 & +0.3 \nl
 N7013 & Field & 1.60 & 3.81 & 3.00 & 6 & +0.2 & 1.08 & 2 & +0.5 \nl
 N7280 & Field & 2.61 & 3.57 & 3.10 & $< 2 $ & +0.7 & 0.07 & $< 2 $
  & +0.7 \nl
 N7332 & Field & 2.12 &  3.72 & 2.86 & 3 & +0.3 & 0.25 & 2 & +0.4 \nl
 N7457 & Field & 1.96 &  2.82 & 2.46 & 8 & -0.2 & 0.46 & 4 & 0 \nl
 N7743 & Field & 2.21 & 3.21 & 2.26 & $< 2 $ & +0.7 & 6.51 & $ < 2$
  & -- \nl
 U11920 & Field & 1.60 & 4.44 & 3.12  & 9 & +0.3 & 0.76 & 3 & +0.7 \nl
\enddata
\tablenotetext{a}{Galaxy ID -- N=NGC, U=UGC}
\tablenotetext{b}{From Guiricin et al. 2000}
\tablenotetext{c}{Estimated with the H$\beta$ index corrected from the
emission through the equivalent width of H$\alpha$ emission line}
\tablenotetext{d}{Estimated with the H$\beta$ index corrected from the
emission through the [OIII]$\lambda$5007 equivalent width}
\tablenotetext{e}{We cannot correct this H$\beta$ index from the emission 
 through H$\alpha$ equivalent width}
\end{deluxetable}

\clearpage

\begin{deluxetable}{llcccccccc}
\tabletypesize{\scriptsize}
\rotate
\tablenum{5}
\tablewidth{22cm}
\tablecaption{Indices and ages for the bulges of the S0 galaxies\label{bulges}}
\tablehead{
\colhead{Galaxy\tablenotemark{a}}  &  
\colhead{Environment\tablenotemark{b}} &
\colhead{H$\beta$}  &
\colhead{Mgb}  &
\colhead{$\langle \mbox{Fe} \rangle$} &
\colhead{T\tablenotemark{c}, Gyr} &
\colhead{[Z/H]\tablenotemark{c}} &
\colhead{EW([OIII]5007), \AA } &
\colhead{$T^{\prime}$\tablenotemark{d}, Gyr} &
\colhead{$\mbox{[Z/H]} ^{\prime}$\tablenotemark{d}}
}
\startdata
 N0080 & Group center & 1.60 &  4.67 & 2.82 & (10) & (+0.2) & 0.08 &
   8 & +0.3 \nl
 N0474 & Group member (pair) & 1.74 & 4.18 & 2.99 & (7) & (+0.2) & 0.35 &
  4 & +0.4 \nl
 N0524 & Group center & 1.07 &  3.65 & 2.00 & $> 15$ & 0? & 0 &
  $> 15$ & 0? \nl
 N0936 & Group center & 1.40 & 4.51 & 2.50 & 15 & 0 & 0.48 & 9 & +0.2 \nl
 N1023 & Group center & 1.43 & 3.94 & 2.69 & (15) & (--0.1) & 0.06 &
  15 & --0.1  \nl
 N1161 & Group member (pair) & 1.71 & 4.24 & 2.86 & (8) & (+0.2) & 0 &
   8 & +0.2 \nl
 N2300 & Group member & 1.50 & 4.85 & 2.84 & (12) & (+0.2) & 0 &
   12 & +0.2 \nl
 N2549 & Group center & 2.22 &  4.05 & 3.15 & 2 & +0.7 & 0.20 &
  2 & +0.7 \nl
 N2655 & Group center & 1.45 &  3.64 & 2.27 & 15 & --0.2 & 0.78 &
  7 & --0.1 \nl
 N2681 & Group center & 2.66 & 2.30 & 1.97 & 3 & --0.2 & 0.42 & 2 &
  --0.2 \nl
 N2685 & Field & 1.41 & 2.63 & 2.37 & ($> 15$) & (--0.3?) & 0.82 & 9 &
  --0.3 \nl
 N2732 & Group member (pair) & 1.62 & 3.53 & 2.30 & 11 & --0.2 & 0.60 &
  6 & 0 \nl
 N2768 & Group center & 1.24 &  4.21 & 2.48 & 13 & 0 & 0.70 & 11 & 0 \nl
 N2787 & Field & 1.03 &  4.54 & 2.38 & $> 15$ & --0.1? & 0.48 &
  $> 15$ & --0.1 \nl
 N2880 & Field & 1.74 &  3.92 & 2.56 & (9) & (0) & 0.12 & 8 & +0.1 \nl
 N2911 & Group center & 0.69 & 3.89 & 2.34 & $> 15$ & ? & 0.78 &
   $> 15$ & $< 0$ \nl
 N2950 & Field & 2.13 & 4.42 & 3.02 & 3 & +0.7 & 0.72 & $< 2$ &
  $> +0.7$ \nl
 N3065 & Group center & 1.54 &  3.94 & 2.51 & 15 & --0.1 & 0.90 &
  4 & +0.2 \nl
 N3098 & Field & 1.96 & 3.57 & 2.33 & (6) & (0) & 0.32 & 4 & 0 \nl
 N3166 & Group member & 2.54 & 3.37 & 2.66 & 2 & +0.3 & 0.35 & $< 2$
   & +0.3 \nl
 N3245 & Group member & 1.70 & 4.30 & 3.02 & (8) & (+0.3) 
      & 0.16 & 5 & +0.4 \nl
 N3384 & Group member & 1.71 & 4.00 & 2.87 & (9) & (+0.2) 
      & 0.18 & 7 & +0.2 \nl
 N3412 & Group member & 2.13 & 3.62 & 2.80 & (3) & (+0.3) 
      & 0.25 & 2 & +0.3 \nl
 N3414 & Group center & 1.08 &  4.70 & 2.47 & 15 & 0 & 1.05 & 9 & +0.2 \nl
 N3607 & Group center & 1.59 &  4.37 & 2.83 & 10 & +0.2 & 0 & 10 & +0.2 \nl
 N3941 & Group center & 1.58 & 3.80 & 2.70 & (13) & (0) 
       & 0.87 & 3 & +0.3 \nl
 N3945 & Group member & 1.38 & 4.19 & 3.12 & (15) & (+0.1) 
       & 0.12 & 14 & +0.1 \nl
 N4026 & UMa cluster & 1.66 & 4.09 & 2.89 & (10) & (+0.2) 
       & 0.59 & 4 & +0.3 \nl
 N4036 & Group center & 0.86 & 3.85 & 2.64 & $> 15$ & ? & 0.36
  & $> 15 $ & $< 0$ \nl
 N4111 & UMa cluster & 1.61 & 3.53 & 2.32 & 11 & --0.2 & 0.34 & 6 & 0 \nl
 N4125 & Group center & 1.61 & 4.64 & 3.13 & 6 & +0.4 & 0.34 & 4 & +0.4 \nl
 N4138 & UMa cluster & 1.03 & 3.40 & 2.06 & 8 & --0.2 & 1.7 & 6 & --0.1 \nl
 N4150 & Group member & 2.22 & 2.51 & 1.87 & 5 & --0.3 & 0.70 & 3 & 
   --0.2 \nl
 N4179 & Group center & 1.69 & 4.35 & 3.04 & (8) & (+0.3) & 0 
       & 8 & +0.3 \nl
 N4233 & Virgo cluster & 1.63 & 4.19 & 2.78 & 11 & +0.1 & 0 & 11 & +0.1 \nl
 N4350 & Virgo cluster & 1.63 & 4.73 & 2.75 & (9) & (+0.3) & 0 & 9 
       & +0.3 \nl
 N4379 & Virgo cluster & 1.66 & 3.98 & 2.29 & (12) & (--0.1) & 0.35 
       & 8 & 0 \nl
 N4429 & Virgo cluster & 1.43 & 4.52 & 2.69 & 15 & +0.1 & 0.19 & 12 &
   +0.1 \nl
 N4526 & Virgo cluster & 1.30 & 4.45 & 2.76 & 3 & +0.6 & 0.06 &
  $> 15 $ & +0.1 \nl
 N4550 & Virgo cluster & 1.66 & 3.14 & 1.95 & 15 & --0.3 & 0.70 &
   6 & --0.2 \nl
 N4570 & Virgo cluster & 1.64 & 4.74 & 2.91 & (8) & (+0.3) & 0 
     & 8 & +0.3 \nl
 N4638 & Virgo cluster & 2.13 & 4.17 & 3.20 & (3) & (+0.7) & 0 
     & 3 & +0.7 \nl
 N4866 & Group member (pair) & 1.50 & 4.08 & 2.23 & (15) & (--0.2) 
    & 0.60 & 8 & 0  \nl
 N5308 & Group member & 1.62 & 4.94 & 3.04 & (7) & (+0.4) & 0.08 & 6 &
  +0.5 \nl
 N5422 & Group member & 1.62 & 4.78 & 3.16 & (7) & (+0.4) & 0.73 &
  3 & +0.7 \nl
 N5574 & Group member & 2.38 & 3.15 & 2.43 & (3) & (+0.2) & 0.36 &
  2 & +0.2 \nl
 N5866 & Group member & 1.73 & 3.62 & 2.95 & 7 & +0.1 & 0.28 &
  5 & +0.2 \nl
 N6340 & Group center & 1.05 & 3.12 & 2.11 & $> 15$ & --0.3 & 0.62
  & $ > 15$ & --0.3 \nl
 N6548 & Group member (pair) & 1.91 & 4.71 & 3.02 & (4) & (+0.5) & 0 &
  4 & +0.5 \nl
 N6654 & Group member (pair) & 1.36 & 4.19 & 2.68 & ($\ge 15$) & (0) 
    & 0.24 & 15 & 0 \nl
 N6703 & Field & 1.92 & 4.45 & 3.16 & 3 & +0.6 & 0.38 & 3 & +0.7 \nl
 N7013 & Field & 2.09 & 3.30 & 2.35 & 3 & +0.1 & 0.60 & 2 & +0.2 \nl
 N7280 & Field & 1.87 & 3.01 & 2.72 & 7 & --0.1 & 0.07 & 7 & --0.1 \nl
 N7332 & Field & 1.60 &  2.66 & 2.14 & (15) & (--0.3) & 0.58 
       & 7 & --0.2 \nl
 N7457 & Field & 2.24 &  3.18 & 2.25 & (4) & (--0.1) & 0.20 & 4 & --0.1 \nl
 N7743 & Field & 2.18 & 3.00 & 2.47 & (4) & (0) & 1.48 & $< 2$ & +0.3 \nl
 U11920 & Field & 1.78 & 4.06 & 2.80  & (8) & (+0.2) & 0.73 & 2 & +0.5 \nl
\enddata
\tablenotetext{a}{Galaxy ID -- N=NGC, U=UGC}
\tablenotetext{b}{From Guiricin et al. 2000}
\tablenotetext{c}{Estimated with the H$\beta$ index corrected from the
emission through the equivalent width of H$\alpha$ emission line}
\tablenotetext{d}{Estimated with the H$\beta$ index corrected from the
emission through the [OIII]$\lambda$5007 equivalent width}
\tablecomments{The values of age and metallicity taken in parentheses
are obtained without correcting the H$\beta$ indices from emission}
\end{deluxetable}

\clearpage

\section{Stellar population properties in the nuclei and the bulges of S0s}

Tables~\ref{nuclei} and \ref{bulges} contain the measured Lick indices 
H$\beta$, Mgb, and
$\langle \mbox{Fe} \rangle \equiv (\mbox{Fe}5270+\mbox{Fe}5335)/2$
for the nuclei and for the bulges correspondingly, as well as the parameters
of the stellar population -- luminosity-weighted age and metallicity --
determined with these indices as described below.
Some galaxies have measurements only for the nuclei or only for the
bulges due to various reasons -- for example, in NGC~5866 the nucleus
is completely obscured by dust and in NGC~676 the bulge measurements
are severely contaminated by a bright star projected at $5\arcsec$
from the nucleus. For the indices presented here, there are 
models based on evolutionary synthesis of simple (one-age,
one-metallicity) stellar populations -- see e.g. \citet{worth94}.
These models allow to estimate the luminosity-weighted mean metallicities
and the ages of the stellar populations by confronting the hydrogen-line
index H$\beta$ to any metal-line index. We are also going to consider the
duration of the last major star-forming episode by confronting
$\langle \mbox{Fe} \rangle$ to Mgb. Chemical evolution models, see 
e.g. \citet{matteuc94}, show  that because of the difference in the timescales
of iron and magnesium production by a stellar generation, the solar
Mg/Fe abundance ratio can be obtained only by very continuous star
formation, and brief star formation bursts, with $\tau \le 0.1$ Gyr, would give 
significant
magnesium overabundance, up to $\mbox{[Mg/Fe]}=+0.3 - +0.4$. In this
work we use recent models by \citet{thomod} because these models
are calculated for several values of [Mg/Fe]: they allow to estimate
Mg/Fe ratios of the stellar populations from Mgb and
$\langle \mbox{Fe} \rangle \equiv (\mbox{Fe}5270+\mbox{Fe}5335)/2$ 
measurements.

Figure~2 presents the $\langle \mbox{Fe} \rangle$ {\bf vs} Mgb diagrams for 
the bulges and Fig.~3 -- the similar diagrams for the nuclei, for all four types 
of environments. For some galaxies
(e.g. NGC 2655 and NGC 2911) where the \ion{N}{1}$\lambda$5199 emission
 is significant, the Mgb indices are corrected from this emission line
according to the prescription of \citet{ge96}. The model
sequences for $\mbox{[Mg/Fe]}=0.0,\, +0.3,\,$ and $+0.5$ are well separated
on the diagrams $\langle \mbox{Fe} \rangle$ {\bf vs} Mgb, so we can
estimate the mean Mg/Fe ratios `by eye'. Surprisingly, the bulges
of the group central galaxies differ from those of the second-ranked
group members: the former have the mean $\mbox{[Mg/Fe]}\approx +0.2$, and
the latter -- $+0.1$.  As by the definition the second-rank group
galaxies are less luminous than the central ones, this difference may
be attributed not to the environment density, but to the galaxy mass effect,
at the first glance. To check this, in Fig.~4 we have plotted the bulges
only for the galaxies within the narrow stellar velocity dispersion range,
$\sigma _*=145-215$ km/s -- in this $\sigma _*$ range the central and 
second-rank group members of our sample have the same {\it mean} 
$\sigma _*$ of 172 km/s; still the difference between the central group 
galaxies and the second-rank members persists in Fig.~4. This tendency
of the S0s in the centers of groups to resemble more the cluster lenticulars, and 
of the second-rank group members and the paired galaxies to be like the field
S0s, is in general confirmed by the nuclei distribution in the 
$\langle \mbox{Fe} \rangle$ {\bf vs} Mgb diagrams (Fig.~3), though there are 
more `outliers' among the nuclei: evidently, the evolution of nuclear stellar
 populations bears more individual features than that of the bulges. 

To break the age-metallicity degeneracy and to determine simultaneously
the mean luminosity-weighted ages and the metallicities of the stellar populations,
we confront the H$\beta$ indices to the combined metal-line index
[MgFe]$\equiv (\mbox{Mgb} \langle \mbox{Fe} \rangle)^{1/2}$ -- by plotting
our data together with the models of \citet{thomod}; earlier we have assured 
that this diagram is insensitive to the Mg/Fe ratio. However we have one serious 
problem here: the absorption-line index H$\beta$ may be contaminated by 
emission, especially in the nuclear spectra. To correct from the emission the
H$\beta$ indices which we have measured we have used data on equivalent 
widths of H$\alpha$ emission lines because H$\alpha$ emission lines are 
always much stronger than H$\beta$ emission lines and because an 
H$\alpha$ absorption line is not deeper than an H$\beta$ absorption line 
in spectra of stellar populations of any age while in intermediate-age population 
spectra  it is much shallower. The emission-line intensity ratio, 
$\mbox{H}\alpha /\mbox{H}\beta$, has been studied well both empirically 
and theoretically. The minimum value of this ratio, 2.5, is known for the case 
of radiative excitation by young stars \citep{burgess}. For other types of 
excitation this ratio is higher. We have no pure \ion{H}{2}-type nuclei in our 
sample,  so here we use the formula 
$EW(\mbox{H}\beta _{\rm emis})=0.25 EW(\mbox{H} \alpha _{\rm emis})$: 
this mean relation is obtained by \citet{sts2001} for a quite 
heterogeneous sample of nearby emission-line galaxies. The data on
$EW(\mbox{H}\alpha _{\rm emis})$ for the nuclei we take mainly from 
\citet{hofil3}.
The bulge H$\beta$ indices were corrected from the emission by using H$\alpha$
equivalent widths obtained with the red MPFS spectra for about a half of the 
sample (28 objects, see the Table~\ref{sample}). Unlike \citet{hofil3} who
obtained $EW(\mbox{H} \alpha _{\rm emis})$ by subtracting a pure 
absorption-line template from the observed spectra, we applied 
a multicomponent Gauss-analysis
to the combinations of the $\mbox{H} \alpha$ absorption and emission lines
which was effective due to mostly different velocity dispersions of stars
and gas clouds in the galaxies under consideration. From the rest, 
16 galaxies have negligible emission lines in the bulge spectra 
($EW(\mbox{[\ion{O}{3}]}) \le 0.3$\AA,
see the Table~\ref{bulges}), and for the others the age estimates 
obtained by using the H$\beta$ indices 'corrected through the H$\alpha$' 
(Table~\ref{bulges}) are indeed only upper limits. To correct in some
way ALL the bulge spectra, we have used also the wide-known approach
which involves the [\ion{O}{3}]$\lambda$5007 emission line; \citet{trager00}
recommend to use the statistical correction 
$\Delta \mbox{H}\beta = 0.6 EW(\mbox{[OIII]}\lambda 5007)$ though they
note that individual ratios H$\beta$/[\ion{O}{3}] may vary between 0.33 and
1.25 within their sample of elliptical galaxies. In Fig.~5 we compare the 
corrections obtained by two different ways for the nuclei. If we
exclude two galaxies with extremely strong emission in the centers
-- NGC 4138 and NGC 7743 -- statistically the two types of the
corrections are indistinguishable; however the accuracy of [\ion{O}{3}]
measuring is not very high due to strong underlying absorption lines
of \ion{Ti}{1}, and the weak emission lines [\ion{O}{3}] with the
equivalent widths of $EW \le 0.3$\AA\ are evidently artifacts. To summarize
this analysis, we conclude that while for mutual comparisons of the age
distributions we must take only the age estimates corrected through the 
[\ion{O}{3}] because this correction can be made for all galaxies of 
the sample, for the individual galaxies having the red spectra
the estimates made with the H$\beta$ indices corrected through the
H$\alpha$ are more reliable due to the facts that the H$\alpha$ emission
is stronger and that the ratio of the Balmer emission lines depends
only on the excitation mechanism unlike the ratio of H$\beta$ to [\ion{O}{3}]
which depends also on the metallicity of the gas.

Figure~6  presents the diagrams  H$\beta$ {\bf vs} [MgFe] for 
the nuclei (top) and for the bulges (bottom) of the galaxies of all types 
of environments with their H$\beta$ indices corrected through the H$\alpha$ to 
the left and with their H$\beta$ indices corrected through the [\ion{O}{3}] to the 
right, correspondingly. By inspecting these diagrams, we determine the ages 
and the metallicities `by eye' that provides an accuracy of $\sim 0.1$ dex in
metallicity and 1~Gyr for the ages less than 8~Gyr and $\sim 2$ Gyr for
older stellar systems which match our accuracy of the Lick
indices. Directly in the diagrams one can see that the range of  the
ages of the nuclei is very wide: they may be as young as 1 Gyr old and as old 
as 15 Gyr old. The bulges are on average older than the nuclei, and in the 
bottom plots one can see a segregation of the galaxies according to their type 
of environment: most the bulges of the group centers and the cluster galaxies 
are older than 5 Gyr, whereas some of the group members and the field lenticulars 
have the bulges as young as 2-3 Gyr old. The metallicity ranges seem to be
similar for the bulges in all types of environments: their [Z/H] are confined
between $\sim -0.3$ and $\sim +0.4$. By fitting formally the metallicity
distributions by Gaussians, we obtain the mean metallicity for the bulges
in dense environments to be  --0.04 and that for the bulges in sparse environments
to be  --0.13,  with the similar rms of 0.5 dex. The nuclei seem to be on average 
more  metal-rich: only three nuclei in the galaxies of sparse environments have the 
metallicity less than the solar.

\citet{kunt} have already reported the difference between
the stellar population characteristics of the early-type galaxies in the
dense and sparse environments. Their measurements were aperture spectroscopy,
and their samples were the Fornax cluster as an example of dense environments
and galaxies without more than 2 neighbors within the search radius of 1.3 Mpc
as an example of sparse environments -- the latter sample is probably
close to our combination of the field plus paired galaxies. They have found
that the E/S0 galaxies in sparse environments are younger than the E/S0
galaxies in the cluster by 2-3 Gyr -- and our result for the S0s is quite 
the same.
But they have also found the anti-correlation between the age and metallicity,
the younger galaxies in sparse environments being on average more metal-rich 
(by 0.2 dex) than the older galaxies in the cluster; while if we see any
metallicity difference, it should be in opposite sense.

In Fig.~7 we plot cumulative distributions of  the ages: the number of galaxies 
not older than $T$ {\bf versus} $\log T$ (in Gyr). We have united the samples of
the brightest group S0s and the cluster galaxies into a `dense environment' 
sample, and the group second-ranked members and the field S0s -- into a `sparse
environment' sample. The effect of environments is seen both for the nuclei
and for the bulges: in sparse environments the stellar populations are, on 
average, younger. The estimates of the median ages are the following: 
3.7 and 6 Gyr for the nuclei of the galaxies in sparse and dense environments,
correspondingly, and 4.8 and 8.3 Gyr for the bulges.

\section{Discussion}

It is a little bit surprising that according to my results, the `dense'  type of 
environment must be ascribed not only to the clusters but also to the centers 
of groups: the first-ranked and the second-ranked S0 galaxies of the groups 
have very different properties of their central stellar 
population. However, this conclusion is close to the recent finding by 
\citet{hickgroup} that the early-type galaxies of Hickson compact groups 
resemble more the cluster galaxies than the field ones. I think it gives us a 
hint that the dynamical effect of close neighbors may play the main role
in evolution rate, and not the mass of the common dark halo.

\clearpage
\begin{table}
\tablenum{6}
\caption{The mean ages of the bulges within fixed stellar velocity
dispersion ranges\label{agevss}}
\begin{tabular}{l|ccc|ccc|}
\tableline
 & \multicolumn{3}{|c|}{Dense environments} &
\multicolumn{3}{|c|}{Sparse environments} \\
Range of $\sigma _*$, km/s & N$_{gal}$ &
$\langle \mbox{T} \rangle$, Gyr & Its rms 
& N$_{gal}$ & $\langle \mbox{T} \rangle$, Gyr & Its rms \\
\tableline
105--145 & 6 & $6.2\pm 2.2$ & 4.9 & 6 & $4.5\pm 1.0$ & 2.3 \\
145--184 & 8 & $6.5\pm 1.0$ & 2.6 & 9 & $6.6\pm 1.6$ & 4.5 \\
185--225 & 7 & $11.6\pm 1.2$ & 2.8 & 5 & $8.6\pm 1.9$ & 3.9 \\
\tableline
\end{tabular}
\end{table}

\clearpage

Recently some evidences have been published \citep{caldwell,nelan} that
the ages of the stellar populations in early-type galaxies are
correlated with the central stellar velocity dispersion. In our sample,
the galaxies in dense environments are on average more massive than
those in sparse environments so one may suggest that the age difference
found above may be due to the mass difference and not to the
environment influence. To check this effect, I have plotted the bulge
age estimates versus the central stellar velocity dispersion in Fig.~8.
Indeed, the correlation is present implying that the more massive bulges
are older; the slope of the regression $\log T$ {\bf vs} $\log \sigma _{*,0}$
is $1.76\pm 0.65$ for the dense environments and $1.30\pm 0.43$ for the
sparse ones with the correlation coefficients of 0.53 and 0.55,
correspondingly. Following \citet{caldwell}, we have calculated the mean ages
of the bulges within narrow ranges of stellar velocity dispersion
(when the age estimate has only the low limit of 15 Gyr, I have ascribed 
the value of 16 Gyr to it). These estimates are given in Table~\ref{agevss}
 -- please compare them
with those in \citet{caldwell}, 7.4 Gyr, rms 4.2 Gyr, in the range of
$\sigma _* =100-160$ km/s, and 9.9 Gyr, rms 4.2 Gyr, in the range of 
$\sigma _* >160$ km/s. One can see from Table~\ref{agevss} that the
ages of the bulges are the same in different types of environments for
the lower bins of $\sigma _*$, 105--145 and 145-185 km/s; but in the
highest bin, 185--225 km/s, the ages are dramatically different, the
massive bulges in dense environments being much older than the massive
bulges in sparse environments. By inspecting  Fig.~8, we notice
that the separation between the bulges in different types of environments
starts from about $\sigma _* =170$ km/s. Taking 7 galaxies in dense
environments and 7 galaxies in sparse environments with the $\sigma _*$
in the range of 170--215 km/s, we obtain $\langle T \rangle =9.7 \pm 1.3$ Gyr,
rms 3.2 Gyr, for the former and $\langle T \rangle =6.6\pm 1.5$ Gyr, rms
3.7 Gyr, for the latter subsample; so the difference is $3.1 \pm  2.0$ Gyr. 
The application of the Student T-statistics
to this double subsample of the massive bulges shows that the mean age
of the massive bulges in dense environments is larger than the mean
age of the massive bulges in sparse environments with the probability
higher than 0.85 (the hypothesis of $\langle T \rangle _{\mbox{dense}}
\le \langle T \rangle _{\mbox{sparse}}$ is rejected at the significance
level of 0.14).
 
\section{Conclusions}

By considering the stellar population properties in the nuclei and the bulges
of the nearby lenticular galaxies in the various types of environments, I 
have found certain differences between the nuclei and the bulges as well
as between the galaxies in dense and sparse environments. The nuclei
are on average younger than the bulges in any types of environments, and
both the nuclei and the bulges of S0s in sparse environments are younger
than those in dense environments. The results of the consideration 
of the Mg/Fe ratios suggest that the main star formation epoch may be
more brief in the centers of the galaxies in dense environments.

\acknowledgements
I am grateful to the astronomers of the Special Astrophysical Observatory
of RAS V.L. Afanasiev, A.N. Burenkov, V.V.Vlasyuk, S.N. Dodonov,
and A.V. Moiseev for supporting the MPFS observations
at the 6m telescope. The 6m telescope is operated under the financial
support of Science Ministry of Russia (registration number 01-43);
we thank also the Programme Committee of the 6m telescope for allocating
the observational time. During the data analysis we have
used the Lyon-Meudon Extragalactic Database (LEDA) supplied by the
LEDA team at the CRAL-Observatoire de Lyon (France) and the NASA/IPAC
Extragalactic Database (NED) which is operated by the Jet Propulsion
Laboratory, California Institute of Technology, under contract with
the National Aeronautics and Space Administration. The study of the 
young nuclei in lenticular galaxies was supported by the grant of the 
Russian Foundation for Basic Researches no. 01-02-16767.

\clearpage

\begin{figure}
\plotone{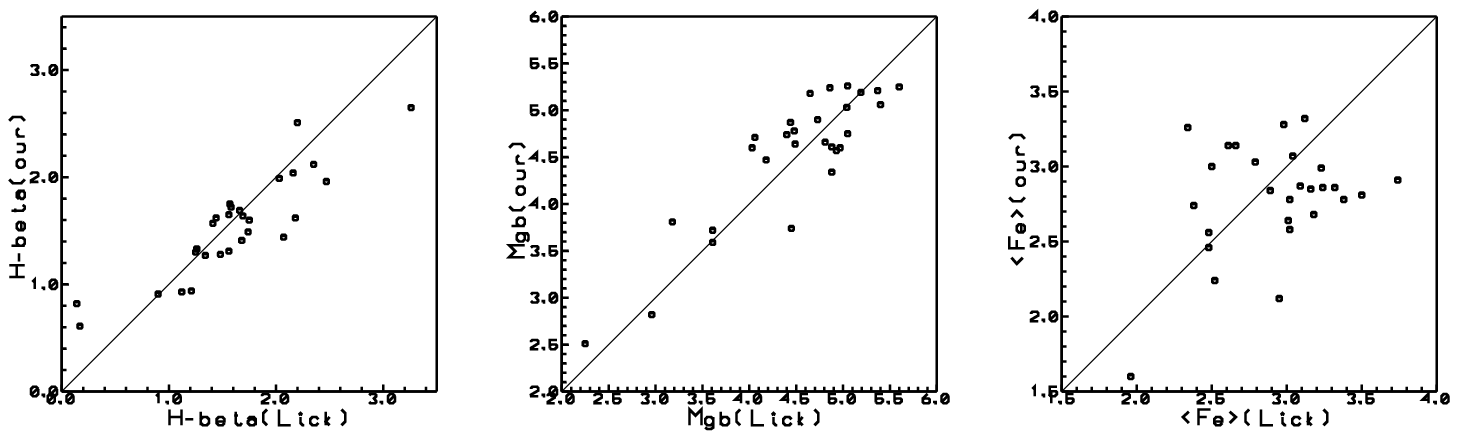}
\caption{The comparison of our measurements of the nuclear Lick indices
with the aperture data of Trager et al. (1998) for 28 common galaxies.
The straight lines are the bissectrices of the quadrants ( 'the lines of equality')}
\end{figure}

\begin{figure}
\plotone{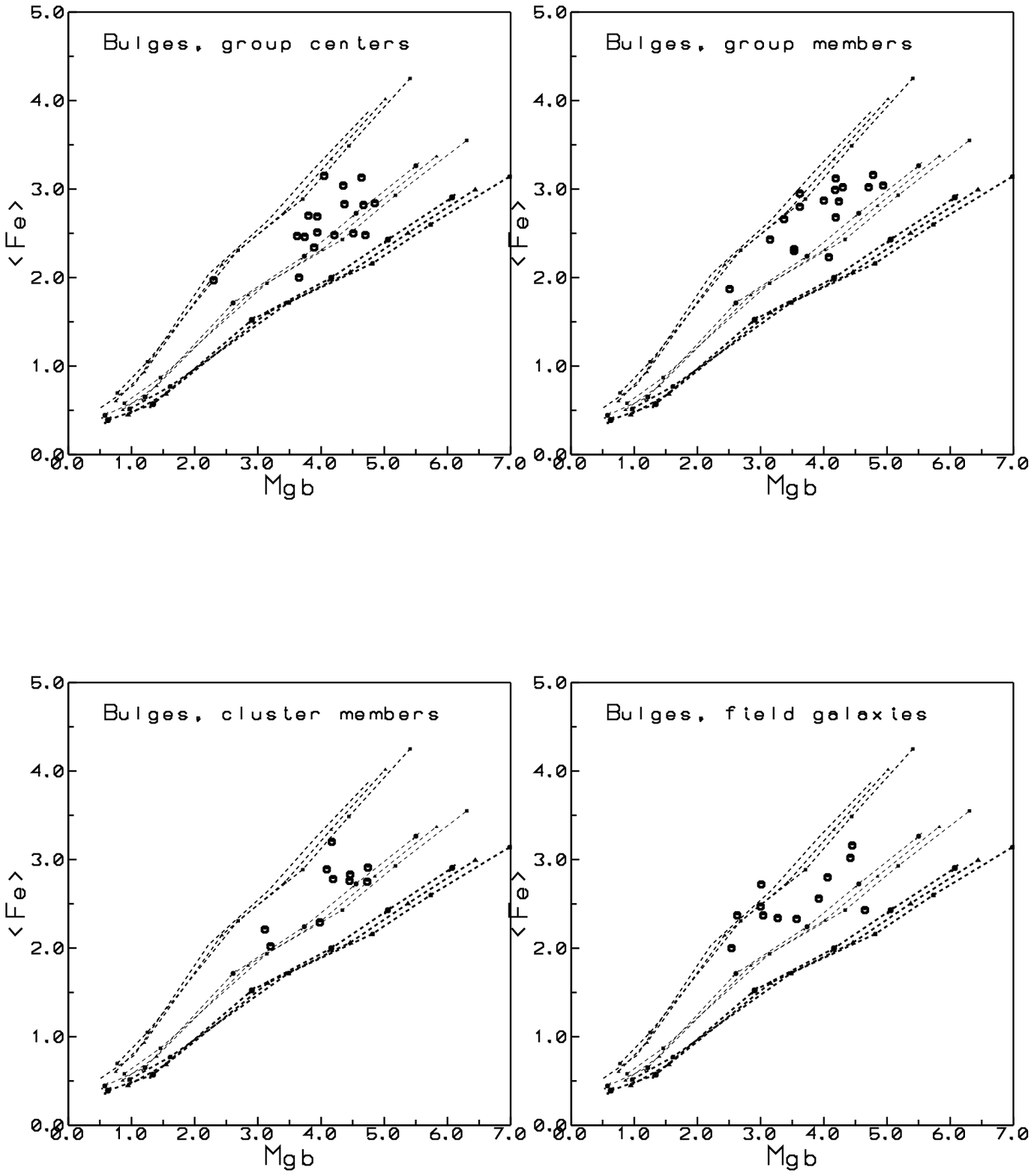}
\caption{The $\langle \mbox{Fe} \rangle$ vs Mgb diagrams for the
bulge index measurements. The typical accuracy of the azimuthally
averaged indices is 0.1~\AA--0.15~\AA. The simple stellar population models 
of Thomas et al.(2003) for three different magnesium-to-iron ratios (0.0, $+0.3$,
and $+0.5$, if the curve triads are taken from top to bottom) and
three different ages (5, 8, and 12 Gyr from top to bottom in every triad) are 
plotted as reference. The small signs along the
model curves mark the metallicities of +0.67, +0.35, 0.00,
--0.33, --1.35, and --2.25, if one takes the signs from
 right to left.}
\end{figure}

\clearpage

\begin{figure}
\plotone{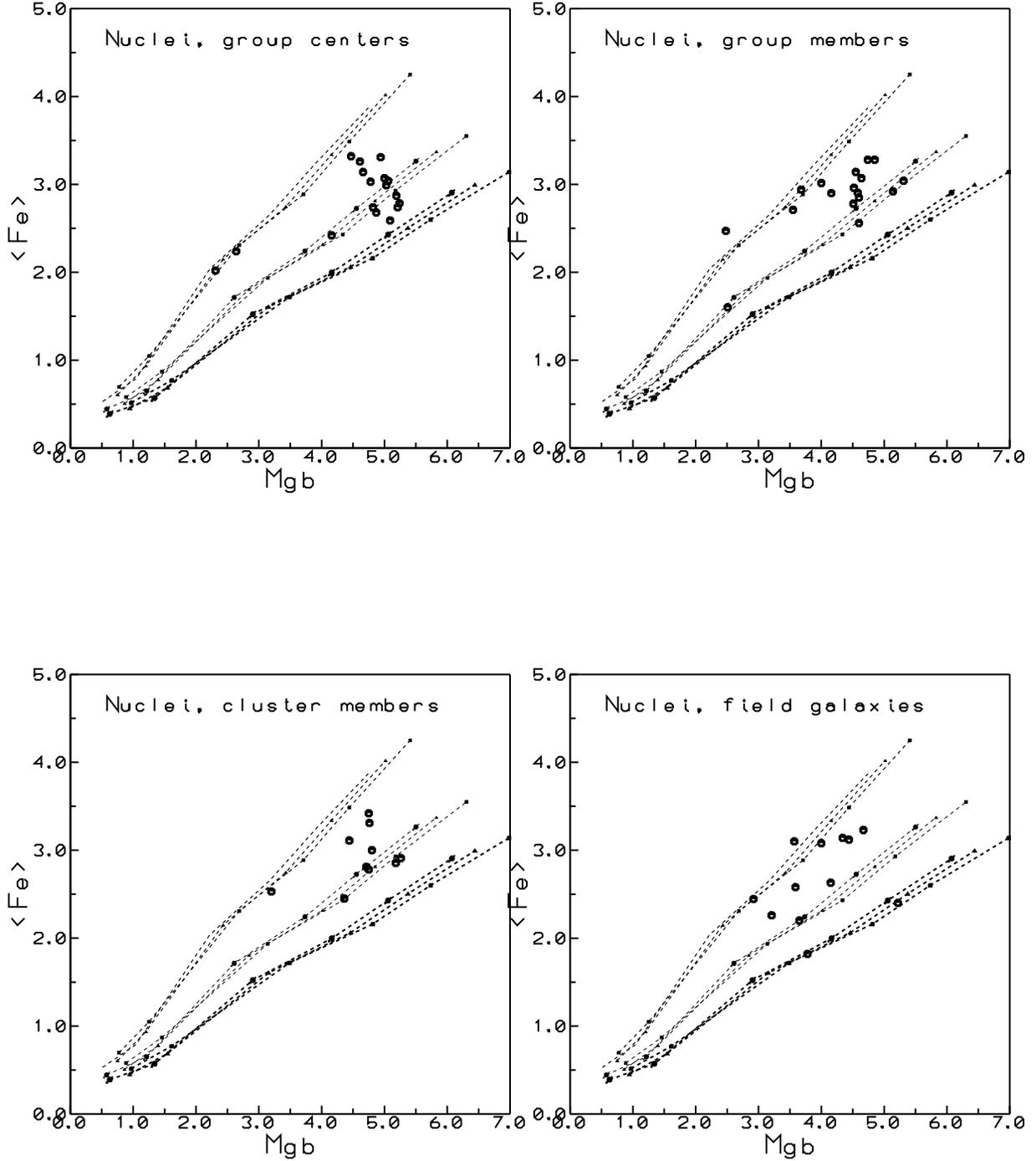}
\caption{The $\langle \mbox{Fe} \rangle$ vs Mgb diagrams for the
nucleus index measurements. The typical accuracy of the nuclear 
indices is 0.1~\AA--0.15~\AA. The simple stellar population models of
Thomas et al.(2003) for three different magnesium-to-iron ratios (0.0, $+0.3$,
and $+0.5$, if the curve triads are taken from top to bottom) and
three different ages (5, 8, and 12 Gyr from top to bottom in every triad) are 
plotted as reference. The small signs along the
model curves mark the metallicities of +0.67, +0.35, 0.00,
--0.33, --1.35, and --2.25, if one takes the signs from
right to left.}
\end{figure}

\clearpage

\begin{figure}
\plotone{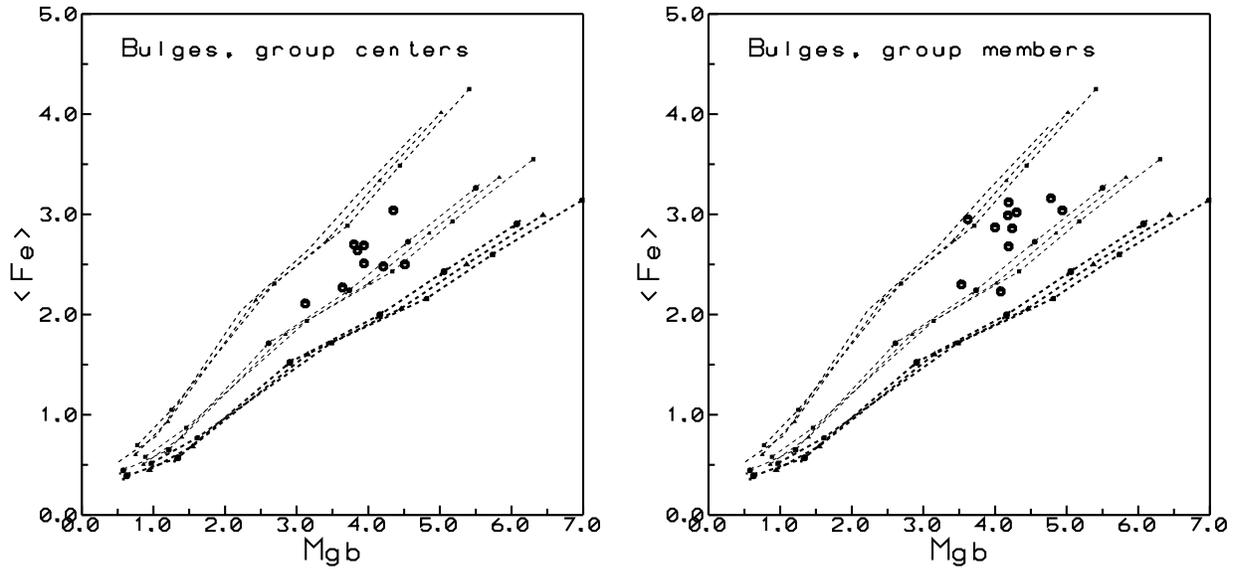}
\caption{The same as in Fig.~2, but only for the group galaxies
with $\sigma _*$ within the range of 145--215 km/s}
\end{figure}

\clearpage

\begin{figure}
\plotone{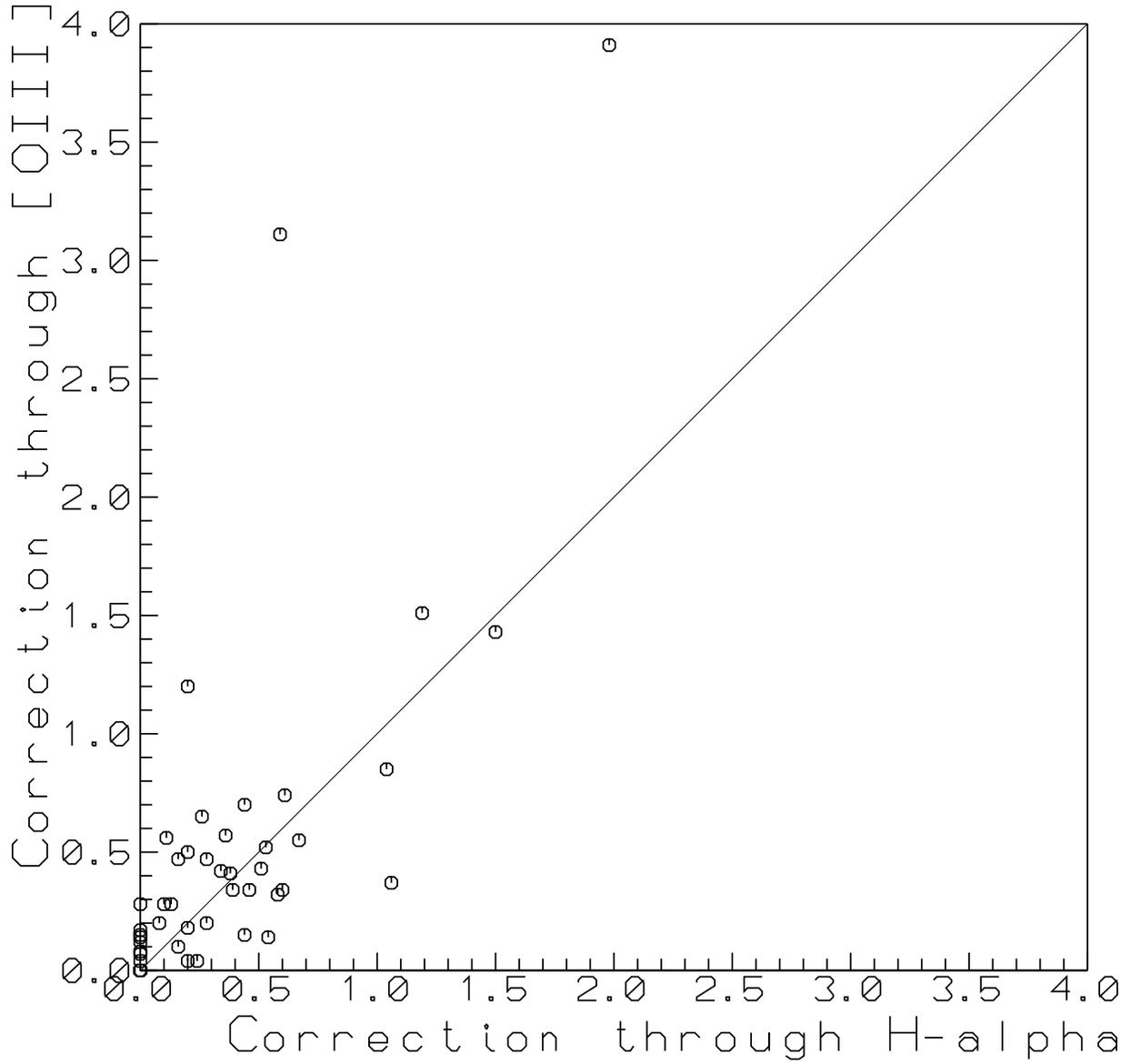}
\caption{The comparison of the H$\beta$ index corrections from the emission
obtained by two different ways -- through H$\alpha$ equivalent widths and
through [\ion{O}{3}] equivalent widths as described in the text. The straight
line is the bissectrice of the quadrant ('the line of equality')}
\end{figure}

\clearpage 

\begin{figure}
\plottwo{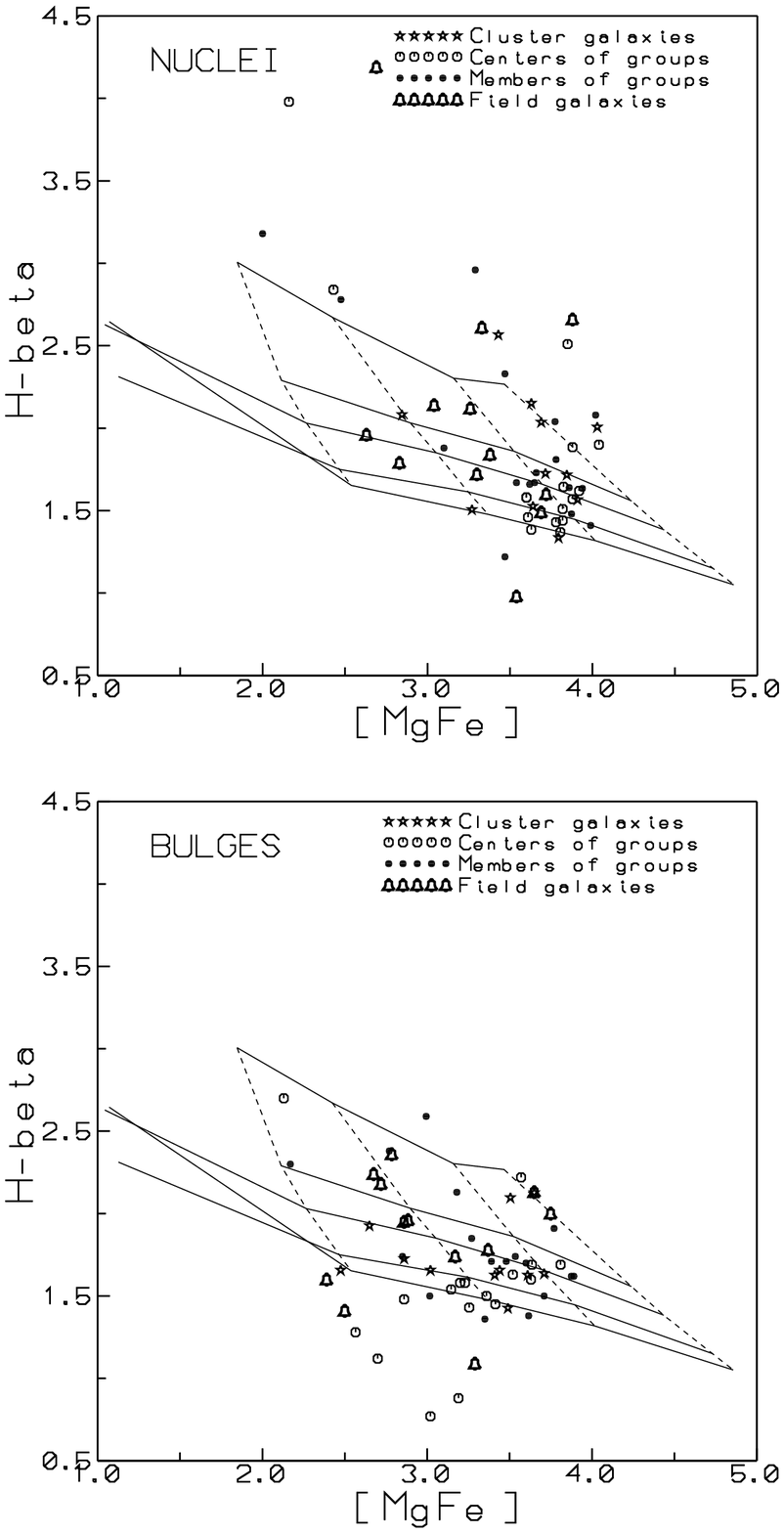}{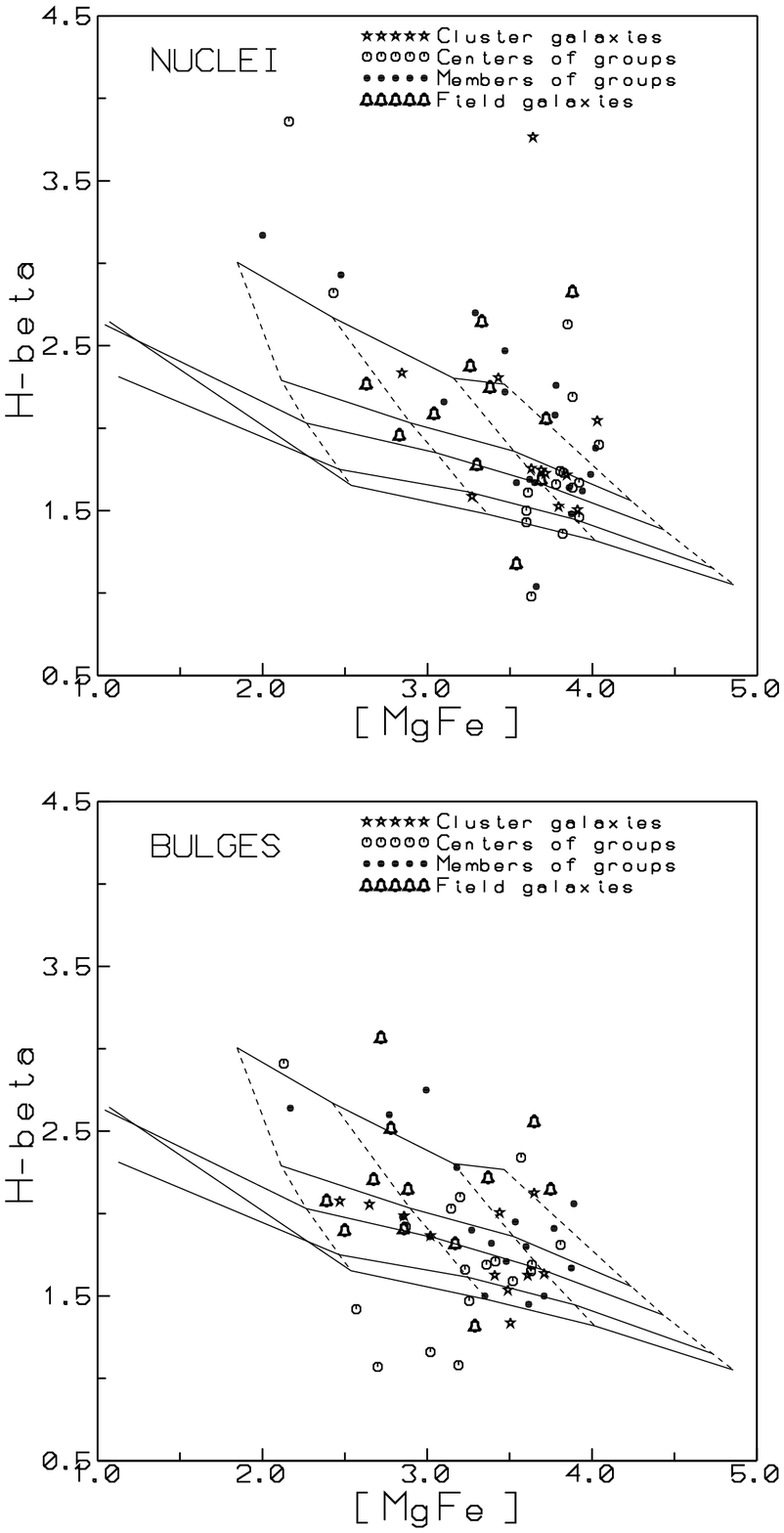}
\caption{The age-diagnostic diagrams for the stellar
populations in the nuclei (\textit{top}) and circumnuclear regions 
(\textit{bottom}) of the galaxies under consideration; 
the H$\beta$-index measurements are corrected
from the emission contamination by using H$\alpha$ in the left plots and
by using [\ion{O}{3}] in the right plots, as described in the text.
The typical accuracy of the indices is 0.1~\AA\ for the combined metal-line index
 and 0.15~\AA\ for the H$\beta$. The stellar population models of 
Thomas et al.(2003) for [Mg/Fe]$=+0.3$ 
and five different ages (2, 5, 8, 12, and 15 Gyr, from top to bottom curves) 
are plotted as reference frame; the dashed lines crossing the
model curves mark the metallicities of +0.67, +0.35, 0.00,
--0.33 from right to left. In the top right plot the nucleus of NGC 7743
which has H$\beta {\rm corr}>6$~\AA\ is omitted.}
\end{figure}

\clearpage

\begin{figure}
\epsscale{0.6}
\plotone{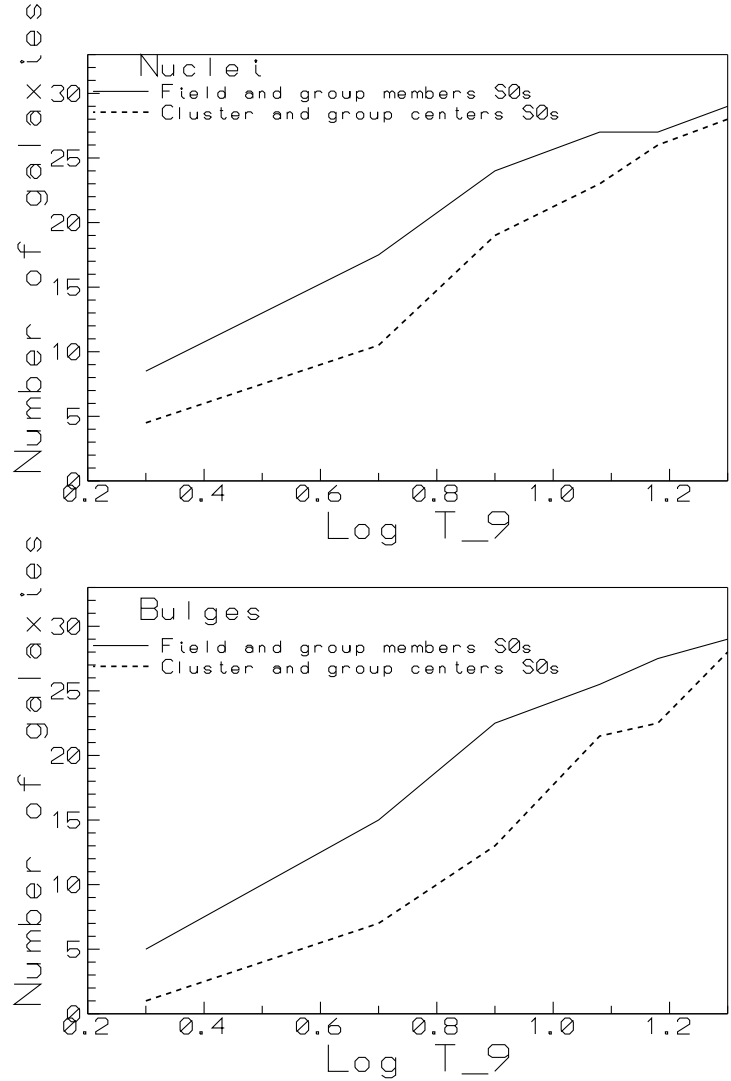}
\caption{Cumulative age distributions: the number of objects
    younger than abcissa which is $\log T$ in Gyr {\bf vs} $\log T$ .
    (\textit{a}) The stellar nuclei of the galaxies
    (\textit{b}) The bulges taken in the rings between
    $R=4^{\prime \prime}$ and $R=7^{\prime \prime}$.}
\end{figure}

\clearpage

\begin{figure}
\plotone{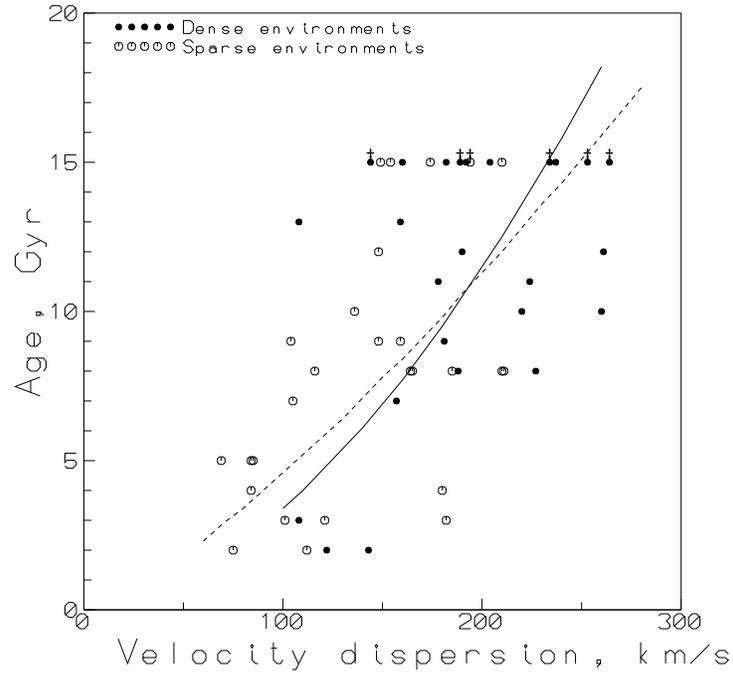}
\caption{Relation between the bulge age estimates obtained
in this work and central stellar velocity dispersions: the regression straight lines
fitting  the dependencies of $\log T$ on $\log \sigma$ are converted into linear
units and plotted by a solid line for the dense environment galaxies and 
by a dashed line for the sparse environment galaxies}
\end{figure}

\end{document}